%% file: skeleton.tex
\documentclass[a4paper,11pt]{article}
\usepackage{pos}

\title{The $\eta_c$-meson leading-twist distribution amplitude}
\ShortTitle{The $\eta_c$ distribution amplitude}

\author[a]{Beno\^{i}t Blossier}
\author[b]{Mariane Mangin-Brinet}
\author[c]{Jos\'{e} Manuel Morgado Ch\'{a}vez}
\author*[a]{Teseo San Jos\'{e}}

\affiliation[a]{
	Laboratoire de Physique des 2 Infinis Irène Joliot-Curie, CNRS/IN2P3,\\
	Université Paris-Saclay, 91405 Orsay Cedex, France}

\affiliation[b]{
	Laboratoire de Physique Subatomique et de Cosmologie, CNRS/IN2P3,\\
	38026 Grenoble, France}

\affiliation[c]{
	D\'{e}partement de Physique Nucl\'{e}aire, Irfu/CEA-Saclay,\\
  	91191 Gif-sur-Yvette Cedex, France}

\renewcommand{\printHeadAuthors}{Teseo San Jos\'{e} et al.}

\emailAdd{san-jose-perez@ijclab.in2p3.fr}
\emailAdd{blossier@ijclab.in2p3.fr}
\emailAdd{mariane@lpsc.in2p3.fr}
\emailAdd{jose-manuel.morgadochavez@cea.fr}

\abstract{In this project, we employ the short-distance factorization to compute the distribution amplitude of the $\eta_c$-meson from Lattice QCD at leading twist. We employ a set of CLS $N_f=2$ ensembles at three lattice spacings and various quark masses to extrapolate the pseudo distribution to the physical point in the isospin limit. We solve the inverse problem modeling the distribution amplitude, and we match our results to the light-cone in the $\msbar$-scheme. We include a complete error budget, and we compare to two alternative approaches: non-relativistic QCD and Dyson-Schwinger equations, finding good agreement with the latter but not with the former.}

\FullConference{The 41st International Symposium on Lattice Field Theory (LATTICE2024)\\
 28 July - 3 August 2024\\
Liverpool, UK\\}

\begin{document}
\maketitle

\section*{Introduction}

Ever since the discovery of hadron structure at \glsxtrshort{slac}, several inclusive and exclusive experimental processes have been discovered that help us with the study of the internal structure of baryons and mesons. From the theoretical side, factorization theorems \cite{Collins:1989gx} are a fundamental tool to understand processes with a large momentum transfer: They divide the cross section in a convolution of hard and soft pieces, where the first can be computed in perturbation theory and the second requires a non-perturbative approach. In this study we focus on the calculation of \glspl{da}, which appear in meson photoproduction $\HepProcess{\Pphoton^*\Pphoton^*\to M}$ and \gls{dvmp} $\HepProcess{\Pphoton^*\Pproton\to M\Pproton}$, and describe the momentum distribution among the quarks of the meson along the longitudinal direction. The more traditional methods to estimate these quantities include \gls{nrqcd}, \gls{ds} equations, light-front dynamics, or light-cone sum rules. The fact that this quantity is defined along the light-cone, see \cref{eq:da}, prevents its direct calculation using \gls{lqcd}. In this study we employ the method of pseudo-distributions \cite{Ji:2013dva,Radyushkin:2017cyf}, which generalizes the definition of \glspl{da} and other functions to space-like separations and provides an algorithm to recover the light-cone physics in a certain limit.

\section*{Methodology}

\begin{figure}
	\centering
	\begin{subfigure}[t]{0.49\textwidth}
		\centering
		\includegraphics[scale=1]{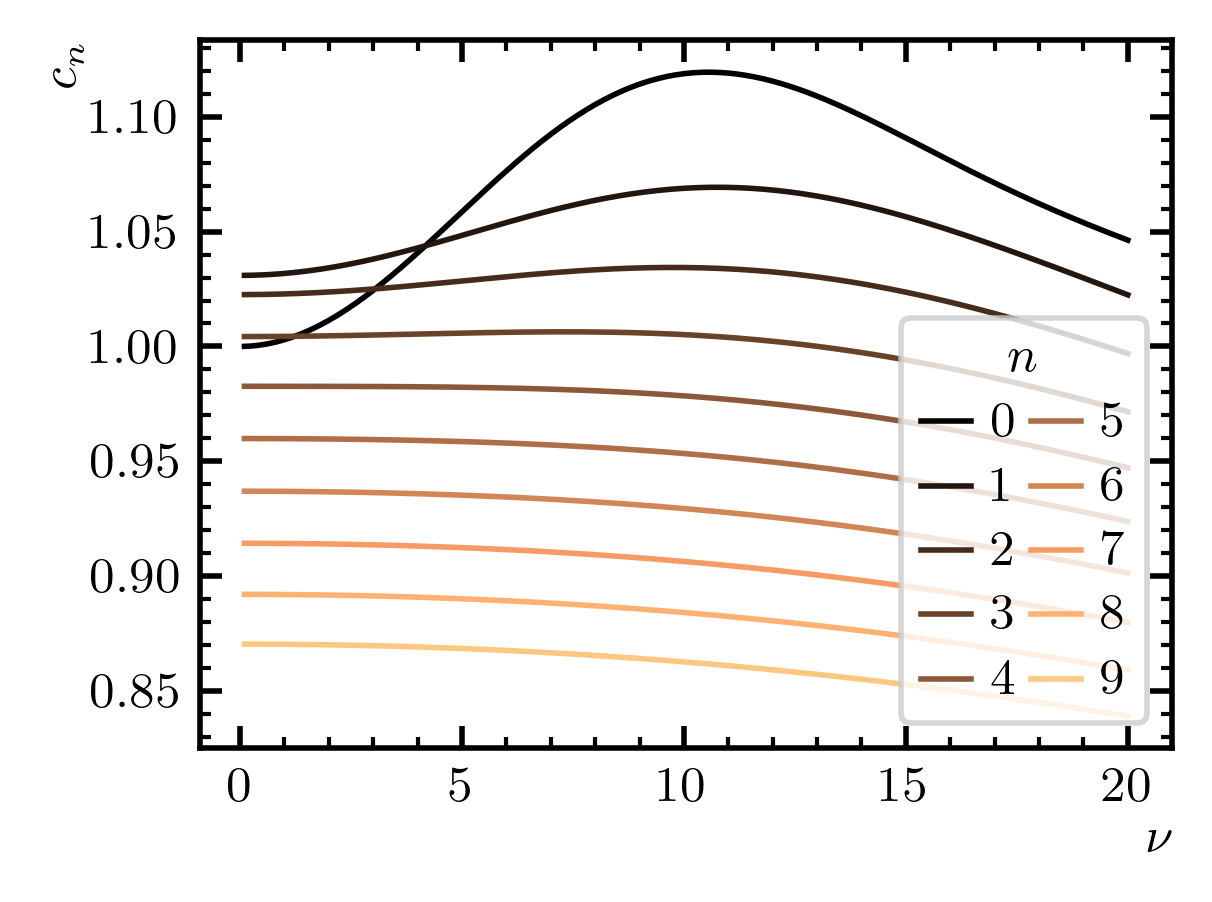}
	\end{subfigure}
	\begin{subfigure}[t]{0.49\textwidth}
		\centering
		\includegraphics[scale=1]{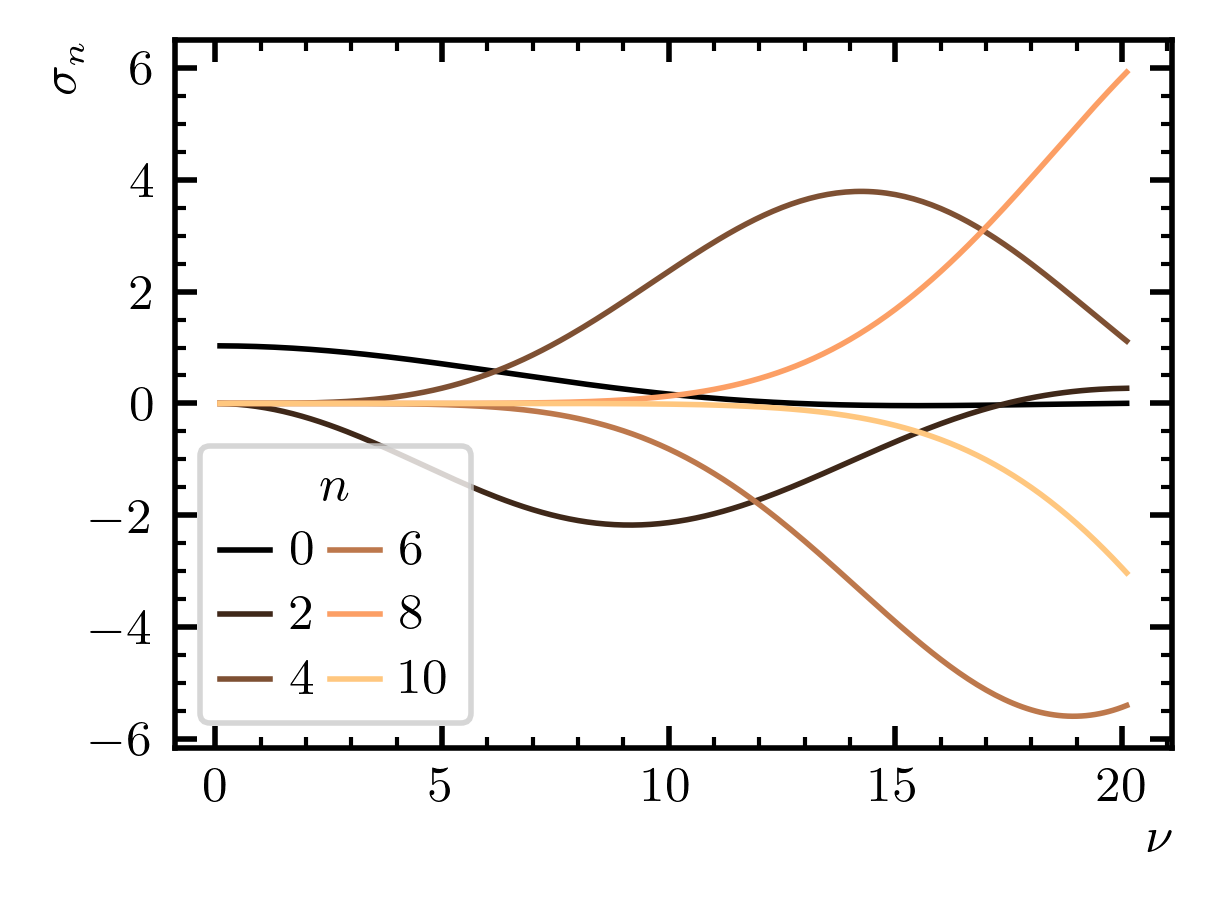}
	\end{subfigure}
	\caption{LEFT: Moments $c_n(\nu,z\mu)$ of the matching kernel. RIGHT: Functions $\sigma_n(\nu,z\mu)$.
	We choose $\lambda=2.7$ and $z_3=5 \times \qty{0.0658}{\femto\meter}$.}
	\label{fig:cn-and-sigma}
\end{figure}

The \gls{da} of charmonium in light-cone coordinates is usually given in terms of the momentum fraction $x$ carried by the quark \cite{Diehl:2003ny}
\begin{equation}
	\label{eq:da}
	\phi(x) = \int \frac{\dd{z^-}}{2\pi} e^{-\iu(x-1/2)p^+ z^-} M^\alpha(p,z),
\end{equation}
where the matrix element itself is the \gls{itda},
\begin{equation}
	\label{eq:itda}
	M^\alpha(p,z) =
	\eval{\mel*{\Petac(p)}{\APcharm(-z/2) \gamma^+ \gamma_5 W(-z/2,z/2) \Pcharm(z/2)}{0}}_{z^+=z^{\text{T}}=0},
\end{equation}
and $\bra*{\Petac}$ is the pseudoscalar meson in the final state, $\ket{0}$ is the \glsxtrshort{qcd} vacuum, $W$ is a Wilson line assuring gauge invariance, $p$ is the hadron momentum, and $\Pcharm$ and $\APcharm$ are the quark fields, which are a distance $z^-$ apart. Unfortunately, the separation along the light cone, $z^2=0$, prevents a direct evaluation in Euclidean space. Instead, we use a generalization of the \gls{itda} in Euclidean metric to space-like separations $z^2 > 0$ \cite{Ji:2013dva,Radyushkin:2017cyf}. To connect this quantity, the \gls{pda}, to the \gls{da} on the light-cone at leading twist we need to take several steps: First, we can separate several higher-twist contributions via a Lorentz decomposition,
\begin{equation}
	\label{eq:lorentz-decomposition}
	M^\alpha(p,z) = 2p^\alpha \mathcal{M}(\nu,z^2) + z^\alpha \mathcal{M}^\prime (\nu,z^2).
\end{equation}
Both terms $\mathcal{M}$ and $\mathcal{M}^\prime$ are Lorentz scalars depending on the Ioffe time $\nu \equiv pz$ and the invariant interval $z^2$. The leading-twist contribution appears in $\mathcal{M}$, and we select it choosing $p=(0,0,p_3,E)$, $z=(0,0,z_3,0)$ and $\alpha=4$. Second, the matrix element $M^\alpha$ renormalizes multiplicatively \cite{Ishikawa:2017faj}, and the renormalization factor depends solely on $z$. We take advantage of the fact that $\nu=0$ is a fixed point of the \gls{itda} to cancel the renormalization factor and avoid its computation entirely. The quantity \cite{Radyushkin:2016hsy,Orginos:2017kos,Karpie:2018zaz}
\begin{equation}
	\label{eq:rpitda}
	\tilde{\phi}(\nu,z^2) \equiv \frac{\mathcal{M}(p,z) \mathcal{M}(0,0)}{\mathcal{M}(0,z) \mathcal{M}(p,0)}
\end{equation}
is known as the \gls{rpitda}, it is \gls{rgi} and $\tilde{\phi}(\nu=0,z)=1$ even at finite lattice spacings. Third, we take the limit $z^2 \to 0$ and match to the light-cone \gls{da} in the $\msbar$-scheme via the relation \cite{Radyushkin:2017lvu,Radyushkin:2019owq}
\begin{equation}
	\label{eq:matching}
	\tilde{\phi}_{\text{lt}}(\nu,z^2) = \int_0^1 \dd{w} C(w,\nu,z\mu)
	\int_0^1 \dd{x} \cos(wx\nu-w\nu/2) \phi_{\text{lt}}(x,\mu)
\end{equation}
where the \gls{da} now depends on the renormalization scale $\mu=\qty{3}{\giga\eV}$, the kernel $C(w,\nu,z\mu)$ takes care of the divergences appearing when $z^2 \to 0$, and the integral in $x$ corresponds to the Fourier transform between $x$ and $\nu$ spaces. In our study we only require the moments of $C(w,\nu,z\mu)$, which are plotted in \cref{fig:cn-and-sigma}. For their expression and more details of the calculation, see \cite{Blossier:2024wyx}.
After introducing the lattice regulator we follow these same steps, and upon arriving to \cref{eq:matching} we encounter an inverse problem, with a finite dataset on the \gls{lhs} to reconstruct a function on the \gls{rhs}. To solve this, we introduce extra information by parameterizing the \gls{da} on the light-cone as
\begin{equation}
	\label{eq:da-model}
	\begin{aligned}
		& \phi_{\text{lt}}(x,\mu) = (1-x)^{\lambda-1/2} x^{\lambda-1/2}
		\sum_{n=0}^\infty d_{2n}^{(\lambda)} \tilde{G}_{2n}^{(\lambda)}(x),
		&
		& d_0^{(\lambda)} = \frac{4^\lambda}{B(1/2,\lambda+1/2)}
	\end{aligned}
\end{equation}
where $B$ is a beta function, $\tilde{G}_{2n}^{(\lambda)}(x)$ are shifted Gegenbauer polynomials defined in the interval $x \in (0,1)$, and the lattice data constraints the coefficients $\lambda$ and $d_{2n}$. A similar approach has been applied to \glspl{pdf} in \cite{Karpie:2021pap}. Setting $\lambda=1.5$, it is possible to recover the conformal expansion of the \gls{da} from \cref{eq:da-model}, and one also obtains the asymptotic result $6x(1-x)$ when $\mu \to \infty$. Indeed, we assume that this expansion, which is true at leading twist and $\order{\alpha_s}$, can also describe non-perturbative data leaving free the coefficient $\lambda$. After all, the series of polynomials form a basis that should be able to describe any smooth function, and even singularities at the endpoints $x=\numlist{0;1}$. Of course, since in our analysis we truncate the series after the first term, this forces the endpoints to be zero for $\lambda > 1/2$. Replacing \cref{eq:da-model} in \cref{eq:matching} simplifies the relation between the fit coefficients and the data,
\begin{equation}
	\label{eq:matching+model}
	\begin{aligned}
		& \tilde{\phi}_{\text{lt}}(\nu,z^2) = \sum_{n=0}^\infty
		\tilde{d}_{2n}^{(\lambda)} \sigma_{2n}^{(\lambda)}(\nu,z^2\mu^2),
		&
		& \tilde{d}_n^{(\lambda)} = \frac{d_n^{(\lambda)}}{4^\lambda},
	\end{aligned}
\end{equation}
where we use a new set of functions $\sigma_{2n}$, plotted in \cref{fig:cn-and-sigma}. Their main feature is that they peak in a certain range of Ioffe times and then vanish. This means that, depending on the domain in Ioffe time of our data, we will be sensitive to more or less of these coefficients (see \cite{Blossier:2024wyx} for more details).

\section*{The lattice calculation}

\begin{figure}
	\centering
	\includegraphics[scale=1]{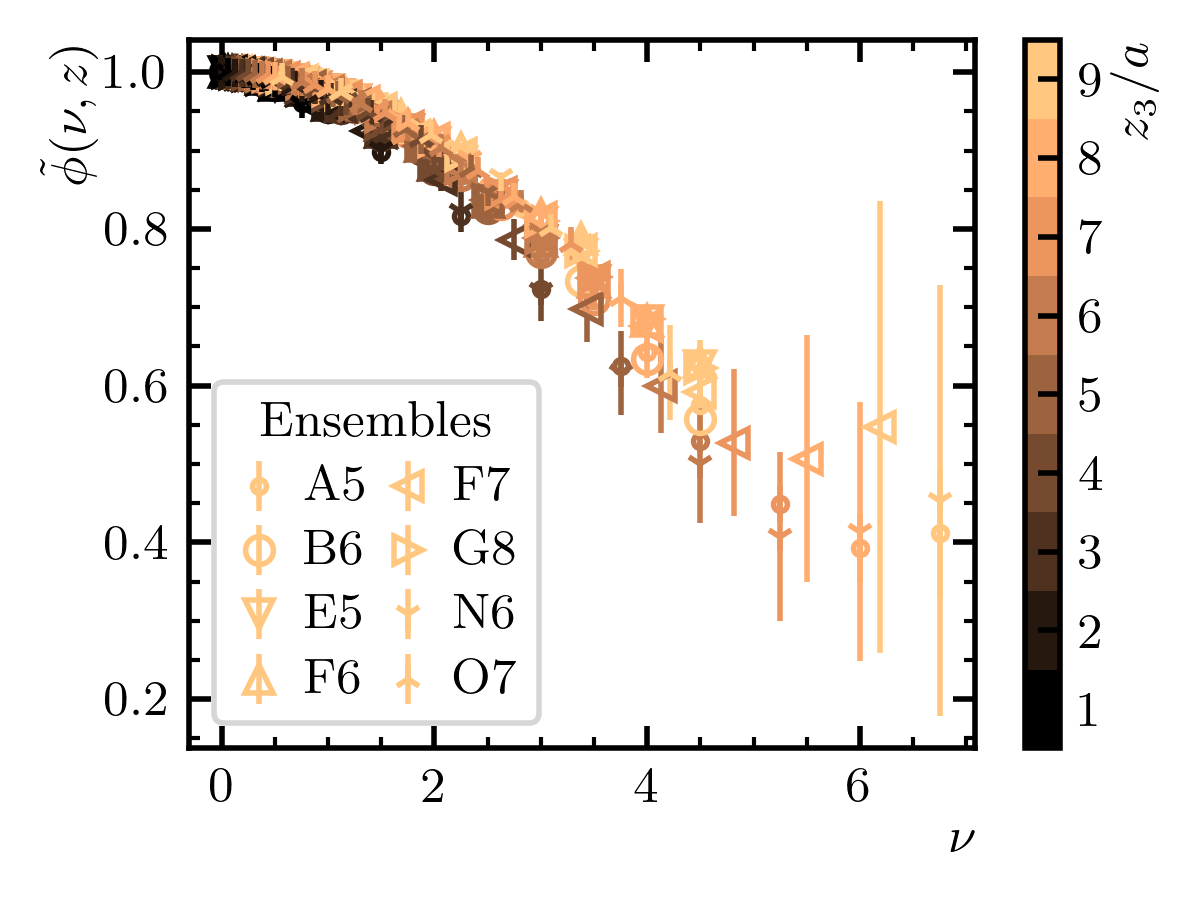}
	\caption{The \gls{rpitda} for the $\Petac$-meson on all ensembles considered in this analysis.}
	\label{fig:rpitda}
\end{figure}

\begin{figure}
	\centering
  	\begin{subfigure}[t]{0.49\textwidth}
		\centering
  		\includegraphics[scale=1]{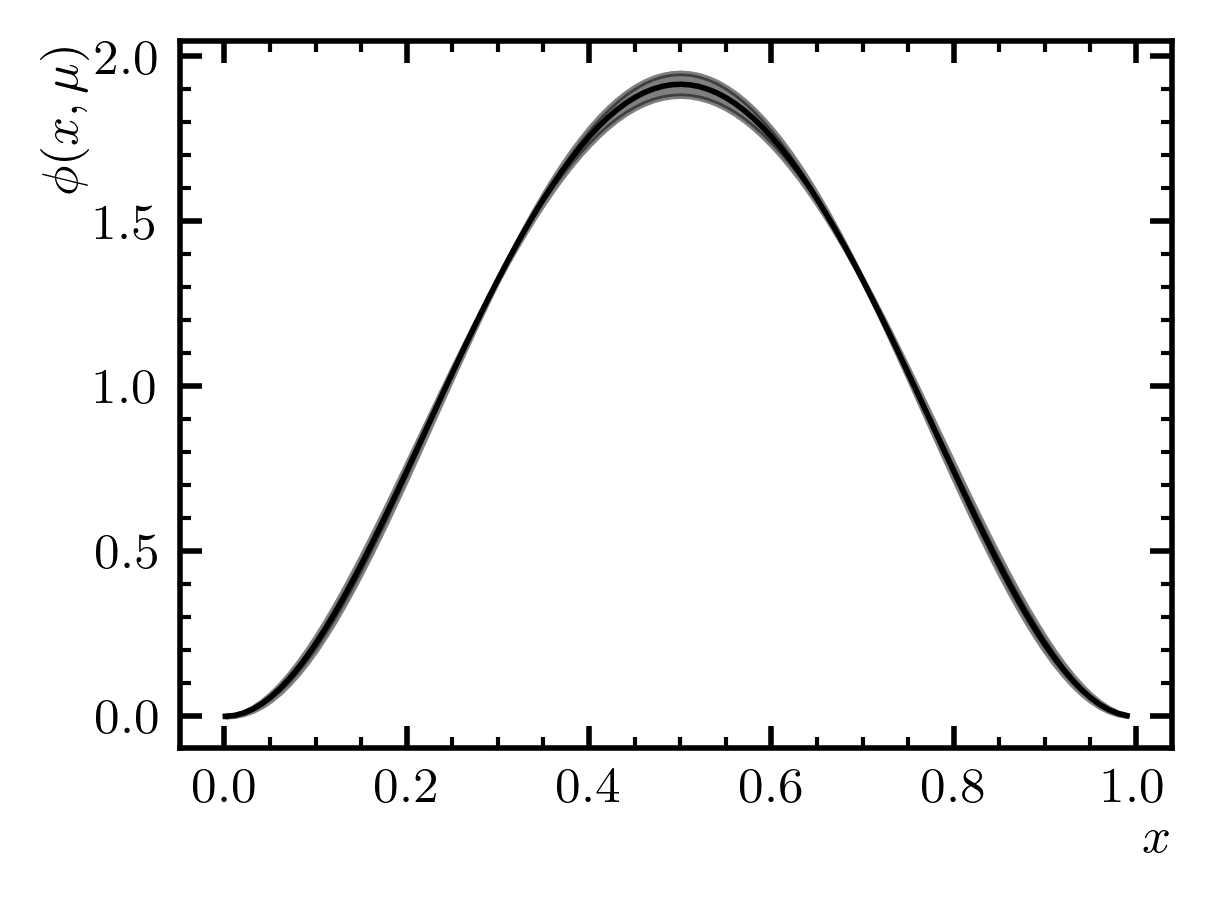}
  	\end{subfigure}
  	\begin{subfigure}[t]{0.49\textwidth}
  		\centering
  		\includegraphics[scale=1]{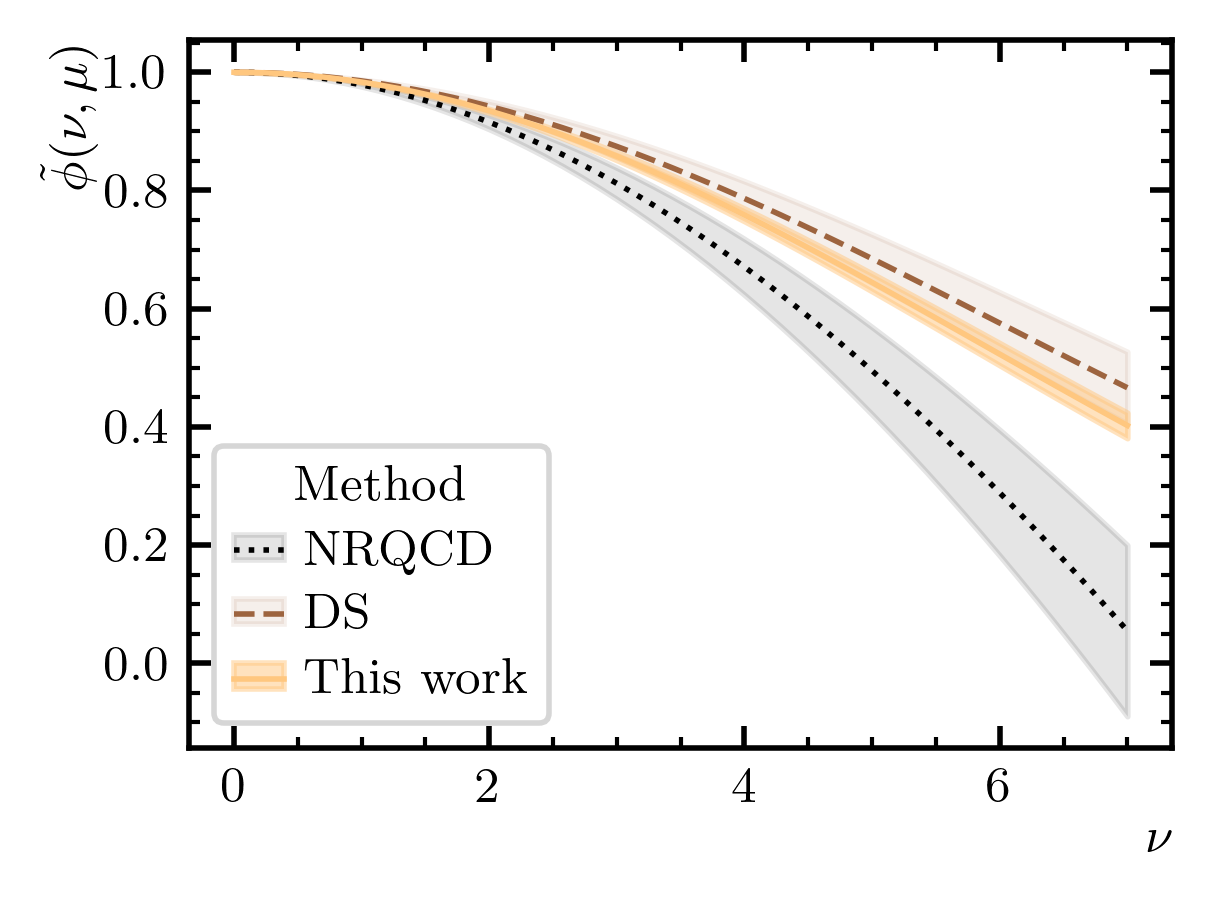}
  	\end{subfigure}
  	\caption{LEFT: The \gls{da} in $x$ space given by \cref{eq:da-lo} using $\lambda = \num{2.73(0.18)}$ at $\mu=\qty{3}{\giga\eV}$. RIGHT: The \gls{da} in Ioffe-time space compared to \gls{nrqcd} \cite{Chung:2019ota} and \gls{ds} equations \cite{Ding:2015rkn}.}
  	\label{fig:da}
\end{figure}

We employ the set of $N_f=2$ \glsxtrshort{cls} ensembles gathered in \cref{tab:ensembles}, which employ the Wilson gauge action and Wilson quarks with non-perturbative $\order{a}$ improvement. The charm-quark mass was tuned so that $m_{\PDs}=m_{\PDs,\text{phys}} = \qty{1968}{\mega\eV}$ \cite{Workman:2022ynf}, while the light-quark masses yield pions in the range $\qty{190}{\mega\eV} < m_{\Ppi} < \qty{440}{\mega\eV}$. We use quark propagators with wall sources diluted in spin, deflated \glsxtrshort{sapgcr} to solve the Dirac equation, and a custom version of the \glsxtrshort{ddhmc} algorithm for the contractions. 
We form the $\Petac$ interpolator solving a $4 \times 4$ \gls{gevp} with different Gaussian smearing levels and the bilinear $\APcharm \gamma_5 \Pcharm$, while its momentum is set via \glspl{ptbc} applied to the charm quark.
\begin{table}
	\centering
	\input{table/cls-ensembles}
	\caption{\glsxtrshort{cls} ensembles included in our analysis. From left to right, we find the labels, the bare couplings and corresponding lattice spacings, the spatial extension ($T=2L$), the pion mass in lattice and physical units, the value of $m_{\Ppi}L$, the bare charm-quark mass and the statistics for each ensemble. An asterisk indicates the ensemble was used only to check for \glspl{fse} and was not used in the extrapolation to the continuum.}
	\label{tab:ensembles}
\end{table}
We only compute the quark-connected Wick contraction of \cref{eq:itda}. We expect a strong \gls{ozi} suppression of the disconnected piece with a factor $\alpha_s^2(\mu) \sim 0.05$. Computing the \gls{rpitda} for all ensembles of \cref{tab:ensembles} yields \cref{fig:rpitda}, where $\tilde{\phi}(\nu,z^2)$ appears as a function of Ioffe time and the color code indicates the extension of the Wilson line. We observe the data fall close to a universal line, with precise data up to $\nu \sim 5$. These points are still affected from a variety of artifacts: We need to take the continuum limit, extrapolate to the physical quark masses, and remove the remaining higher-twist contamination which includes the target-mass corrections. The latter are proportional to $z^2 m_{\Petac,\text{phy}}^2$, which can be sizeable for the $\Petac$-meson.
We fit the lattice data $\tilde{\phi}_e(\nu,z^2)$ to the following model to separate all these effects from the leading-twist \gls{pda} that can be matched to the light-cone using \cref{eq:matching+model},
\begin{equation}
	\label{eq:continuum-extrapolation}
	\begin{aligned}
		\tilde{\phi}_e(\nu,z^2) &= \tilde{\phi}_{\text{lt}}(\nu,z^2)
		+ \frac{a}{\abs{z}} A_1(\nu) + a\Lambda B_1(\nu) + z^2 \Lambda^2 C_1(\nu)
		\\
		& + \frac{a}{\abs{z}}
		\Big(
			\Lambda^{-1} [m_{\Petac}-m_{\Petac,\text{phy}}] D_1(\nu)
			+ \Lambda^{-2} [m_{\Ppi}^2-m_{\Ppi,\text{phy}}^2] E_1(\nu)
		\Big).
	\end{aligned}
\end{equation}
\begin{table}
	\centering
	\input{table/results}
	\caption{Fit parameters in \cref{eq:continuum-extrapolation}. The first column indicates the parameter; the second column the expected value, the statistical and the various systematic uncertainties; the third is the result removing the heaviest pion mass; the fourth keeps only $m_{\Ppi} < \qty{300}{\mega\eV}$ and the fifth checks that the results are stable when we only keep $z_3 < \qty{0.5}{\femto\meter}$. The coefficients $a_{1,2}$, $b_{1,2}$, etc., correspond to the auxiliary functions $A_1$, $B_1$, etc.}
	\label{tab:results}
\end{table}
The leading lattice artifacts are $\order{a}$ because we do not improve the matrix element \cref{eq:lorentz-decomposition}, only the action. The auxiliary functions $A_1$, $B_1$, etc. have a similar form to \cref{eq:matching+model} and have their own fit parameters (their explicit form appears in \cite{Blossier:2024wyx}). The value of $\lambda$ is shared, as it only serves to specify a basis of polynomials. We render all terms dimensionless using $\Lambda \equiv \Lambda_{\text{QCD}}^{(2)} = \qty{330}{\mega\eV}$ \cite{FlavourLatticeAveragingGroupFLAG:2021npn}. Fitting the data shown in \cref{fig:rpitda} to \cref{eq:continuum-extrapolation} yields the results displayed in \cref{tab:results}. We are only sensitive to the first coefficient $d_0^{(\lambda)}$, such that the \gls{da} given in \cref{eq:da-model} is reduced to
\begin{equation}
	\label{eq:da-lo}
	\phi_{\text{lt}}(x,\mu) = \frac{4^\lambda (1-x)^{\lambda-1/2} x^{\lambda-1/2}}{B(1/2,1/2+\lambda)}
\end{equation}
with $\lambda = \num{2.73(0.18)}$ adding in quadrature all uncertainties of \cref{tab:results}. \Cref{eq:da-lo} appears in \cref{fig:da} both in $x$ space and Ioffe-time space. The latter is especially useful to compare to other theoretical determinations because the \gls{da} has to be analytic. In particular, we compare to two alternative approaches, one using the \gls{nrqcd} framework \cite{Chung:2019ota} and another employing \gls{ds} equations \cite{Ding:2015rkn}. We observe good agreement with the latter, which is also a non-perturbative approach, while \gls{nrqcd} differs significantly at larger Ioffe times. Since \gls{nrqcd} relies in a series expansion and the entire $\nu$ dependence is given by the first quantum and relativistic corrections, which become very sizeable for large Ioffe times, we think it is important to know the corrections at next order.

\section*{Conclusions and outlook}

We compute the leading-twist contribution to the \gls{da} of the $\Petac$-meson with a set of $N_f=2$ \glsxtrshort{cls} ensembles using the method of short-distance factorization. Thanks to the various lattice spacings and quark masses we can extrapolate to the physical point in the isospin limit. The \gls{da} on the light-cone is parameterized in \cref{eq:da-lo}. We explore several sources of systematic uncertainty and conduct various crosschecks that show that our results (given in \cref{tab:results}) are stable. The method developed in this work can now be applied to other states, like $J/\psi$, with a more complicated Lorentz structure and bigger impact in the upcoming \glsxtrshort{eic} experiment.

\acknowledgments

The work by T.~San Jos\'{e} is supported by Agence Nationale de la Recherche under the contract ANR-17-CE31-0019. The work by J.~M.~Morgado Ch\'{a}vez has been supported by P2IO LabEx (ANR-10-LABX-0038) in the framework of Investissements d’Avenir (ANR-11-IDEX-0003-01). This project was granted access to the HPC resources of TGCC (2021-A0100502271, 2022-A0120502271 and 2023-A0140502271) by GENCI. The authors thank Michael Fucilla, C\'{e}dric Mezrag, Lech Szymanowski and Samuel Wallon for valuable discussions.

\bibliographystyle{JHEP}
\bibliography{bib}

\end{document}

%% file: table/cls-ensembles.tex
\resizebox{\columnwidth}{!}{
\begin{tabular}{lccc cccc c}
\toprule
  id &
  $\beta$ &
  $\spacing~[\unit{\femto\metre}]$ &
  $L/a$ &
  $am_{\Ppi}$ &
  $m_{\Ppi}~[\unit{\mega\eV}]$ &
  $m_{\Ppi}\boxlen$ &
  $\kappa_{\Pcharm}$ &
  Measurements \\
\midrule
  A5  & 5.2 & 0.0755(9)(7) & $32$ & \num{0.1265(8)} & 331 & 4.0 & 0.12531 & 1980 \\
  B6  &     &              & $48$ & \num{0.1073(7)} & 281 & 5.2 & 0.12529 & 1180 \\
\midrule
  D5* & 5.3 & 0.0658(7)(7) & $24$ & \num{0.1499(1)} & 449 & 3.6 & 0.12724 & 1500 \\
  E5  &     &              & $32$ & \num{0.1458(3)} & 437 & 4.7 & 0.12724 & 2000 \\
  F6  &     &              & $48$ & \num{0.1036(3)} & 311 & 5.0 & 0.12713 & 1200 \\
  F7  &     &              & $48$ & \num{0.0885(3)} & 265 & 4.3 & 0.12713 & 2000 \\
  G8  &     &              & $64$ & \num{0.0617(3)} & 185 & 4.1 & 0.12710 & 1790 \\
\midrule
  N6  & 5.5 & 0.0486(4)(5) & $48$ & \num{0.0838(2)} & 340 & 4.0 & 0.13026 & 1900 \\
  O7  &     &              & $64$ & \num{0.0660(1)} & 268 & 4.2 & 0.13022 & 1640 \\
\bottomrule
\end{tabular}
}

%% file: table/results.tex
\begin{tabular}{crrrr}
\toprule
$\lambda$ & \num{2.73+-0.12+-0.12+-0.06}       & \num{2.75+-0.12}      & \num{2.61+-0.15}      & \num{2.62+-0.10}      \\
$a_{1,2}$ & \num{-7.58+-0.05+-0.59+-0.55}      & \num{-8.12+-0.05}     & \num{-8.68+-0.13}     & \num{-6.76+-0.04} \\
$b_{1,2}$ & \num{0.88+-0.07+-0.08+-0.06}       & \num{0.89+-0.07}      & \num{0.77+-0.10}      & \num{0.81+-0.07}\\
$c_{1,2}$ & \num{-0.042+-0.002+-0.005+-0.001}  & \num{-0.0428+-0.0022} & \num{-0.0440+-0.0028} & \num{-0.0407+-0.0024} \\
$d_{1,2}$ & \num{-2.221+-0.015+-0.063+-0.15}   & \num{-2.368+-0.015}   & \num{-2.52+-0.04}     & \num{-2.000+-0.011}\\
$e_{1,2}$ & \num{-0.0897+-0.001+-0.159+-0.116} & \num{-0.1700+-0.0016} & \num{-0.321+-0.005}   & \num{-0.06848+-0.00012}\\
\bottomrule
\end{tabular}